\documentstyle[12pt,epsf]{article}
\newcommand{\ra}{\rangle}
\newcommand{\la}{\langle}

\newcommand{\eps}{\epsilon}
\newcommand{\II}{{\cal I}}

\newcommand{\NN}{{\cal N}}

\newcommand{\wt}{\widetilde}

\newcommand{\wb}{\bar}

\newcommand{\be}{\begin{equation}}
\newcommand{\ee}{\end{equation}}
\newcommand{\ben}{\begin{eqnarray}\displaystyle}
\newcommand{\een}{\end{eqnarray}}
\newcommand{\refb}[1]{(\ref{#1})}
\newcommand{\p}{\partial}
\newcommand{\sectiono}[1]{\section{#1}\setcounter{equation}{0}}

\begin{document}

{}~ \hfill\vbox{\hbox{hep-th/9805019}\hbox{MRI-PHY/P980545}
}\break

\vskip 3.5cm

\centerline{\large \bf Stable Non-BPS Bound States of BPS
D-branes}

\vspace*{6.0ex}

\centerline{\large \rm Ashoke Sen
\footnote{E-mail: sen@mri.ernet.in}}

\vspace*{1.5ex}

\centerline{\large \it Mehta Research Institute of Mathematics}
 \centerline{\large \it and Mathematical Physics}

\centerline{\large \it  Chhatnag Road, Jhoosi,
Allahabad 211019, INDIA}

\vspace*{4.5ex}

\centerline {\bf Abstract}

S-duality symmetry of type IIB string theory predicts the
existence of a stable non-BPS state on an orbifold five plane of
the type IIB theory if the orbifold group is generated by the
simultaneous action of $(-1)^{F_L}$ and the reversal of sign of
the four coordinates transverse to the orbifold plane. We
calculate the mass of this state by starting from a pair of
D-strings carrying the same charge as this state, and then
identifying the point in the moduli space where this pair
develops a tachyonic mode, signalling the appearance of a bound
state of this configuration into the non-BPS state.

\vfill \eject

\tableofcontents

\baselineskip=18pt

\sectiono{Introduction and Summary} \label{s1}

Quite often duality symmetries in string theory predict the
existence of stable solitonic states in string theory which are
not BPS states, but are stable due to the fact that they are the
lightest states carrying a given set of charge quantum numbers.
Several examples of this kind were discussed in 
ref.\cite{NONBP,BERKOL}.
A class of examples discussed in \cite{NONBP} involved a Dirichlet
$p$-brane\cite{DBRANE} on top of an orientifold
$p$-plane\cite{ORIENT}. This system has an
SO(2) gauge field living on the brane and massive non-BPS states
charged under this SO(2) gauge field. These states must remain
stable in the strong coupling limit since there are no states
into which they can decay. In \cite{NONBP} we identified these
states in the dual weakly coupled theory for
$p=4$, 6 and 7. In this paper we shall focus on the case $p=5$.
\begin{figure}[!ht]
\begin{center}
\epsfbox{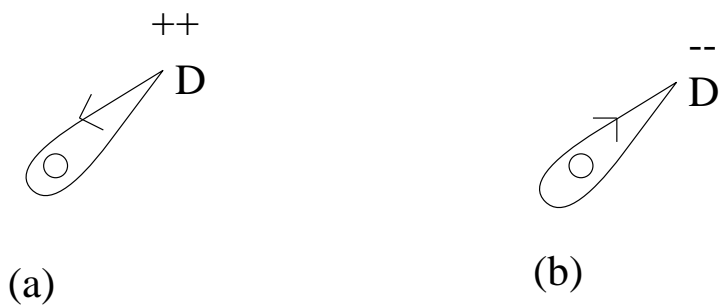}
\end{center}
\caption[]{\small Open string configuration giving charged state
on the world-volume of the D5-brane O5-plane system. The $\circ$
denotes the location of the O5-plane and the D denotes the
location of the D5-brane.} \label{f1}
\end{figure}
\begin{figure}[!ht]
\begin{center}
\epsfbox{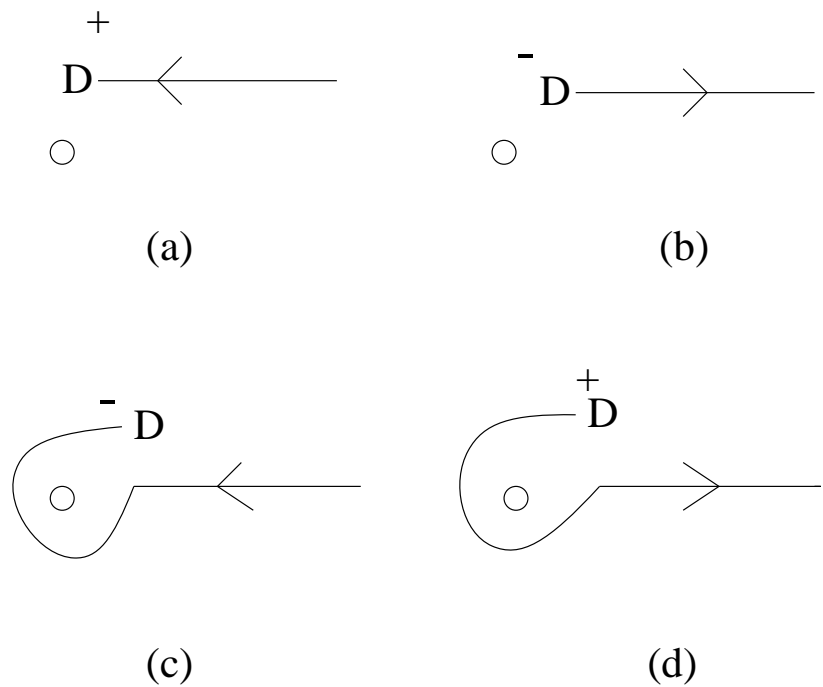}
\end{center}
\caption[]{\small Four possible configurations of fundamental
string ending on a D5-brane $-$ O5-plane system.}
\label{f2}
\end{figure}

Since most of this paper involves details of calculations
involving D-branes, we shall summarise the main results in this
section.
The orientifold 5-plane (O5-plane) in type IIB theory is obtained
by modding out type IIB string theory in ten dimensional
Minkowski space $R^{9,1}$ by $\Omega\cdot \II_4$, where $\Omega$
denotes the world-sheet parity transformation and $\II_4$ denotes
the reversal of sign of four of the space-like coordinates.
For a Dirichlet 5-brane (D5-brane) on top of an 
O5-plane, the state carrying charge under the SO(2)
gauge field on the world volume arises from an open string state
which starts on the D5-brane, goes once around the orientifold
plane, and ends on the D5-brane. Two different orientations of
the string give states of opposite charge on the world-volume, as
shown in Fig.\ref{f1}. 
(Although we are considering a configuration
of coincident D5-O5 system, we shall often display them as a
resolved pair for the sake of clarity.)
We shall normalise the charge in such a way that
the states displayed in Fig.\ref{f1} carry charge $2$ and $-2$
respectively.\footnote{This convention differs from the one used
in ref.\cite{NONBP} by a factor of 2.}
We shall denote these states as ++ and $--$ respectively. 
Besides these non-BPS states, there are other
BPS configurations which carry charge under the world-volume
gauge field. These are semi-infinite
fundamental strings (F-strings) ending on
the D-brane. There are four different configurations of
fundamental strings which differ from each other by the
orientation of the string, and the path followed by the F-string
from infinity to the D5-brane. These four configurations are
shown in Fig.\ref{f2} and carry charges $\pm 1$ under the D-brane
world-volume gauge field.  From the point of view of the D-brane
world-volume field theory these states have
infinite mass due to infinite length of the semi-infinite
F-strings.

In seeking a description of these states in the strong coupling
limit, we go to the dual type IIB theory by making an S-duality
transformation. Since S-duality transforms $\Omega$ to
$(-1)^{F_L}$, where $F_L$ denotes the contribution to the
space-time fermion number from the left-moving sector of the
world-sheet, we expect the O5-plane to get mapped to the fixed
plane of the orbifold $R^{9,1}/(-1)^{F_L}\cdot \II_4$. However,
as was argued in \cite{DUORBI}, the orbifold plane actually
describes the dual of the coincident O5-plane $-$ D5-brane
system. One
way to see this is to note that the twisted sector in the
orbifold theory contains a gauge field living on the orbifold
plane. Since an isolated O5-plane does not have a gauge field
living on the plane, the orbifold plane cannot be dual to an
isolated O5-plane. On the other hand a D5 brane on top of an
O5-plane has precisely the same degrees of freedom as those 
appearing on the orbifold plane.
\begin{figure}[!ht]
\begin{center}
\epsfbox{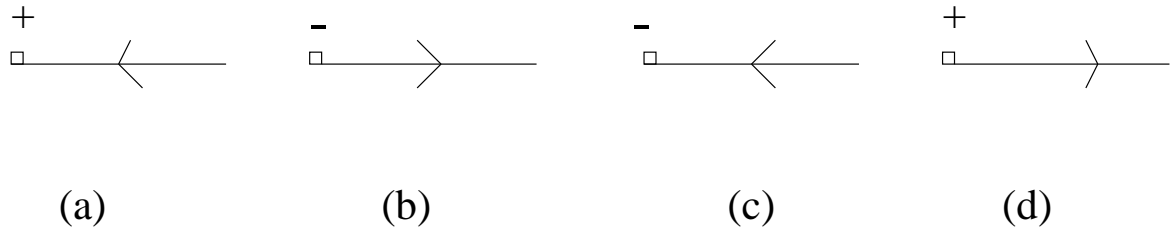}
\end{center}
\caption[]{\small S-dual of the configurations shown in
Fig.\ref{f2}. Here the little square denotes the location of the
orbifold plane.}
\label{f3}
\end{figure}

If this identification is correct, then the orbifold plane must
support configurations which are S-dual to the configurations
shown in Figs.\ref{f1} and \ref{f2}. Configurations dual to
\ref{f1} are the states carrying charge $\pm 2$ units under the
(appropriately normalised) twisted sector gauge
field living on the orbifold plane. We shall refer to these as ++
and $--$ states respectively. On the other hand the dual of
the configurations in Fig.\ref{f2} are D-strings ending on the
orbifold plane, as shown in Fig.\ref{f3}. The ends of these
strings should
carry $\pm 1$ unit of charge under the twisted sector gauge field.
Note that for each
orientation of the D-string, we must find two states, carrying
opposite charges under the twisted sector gauge field. 
\begin{figure}[!ht]
\begin{center}
\epsfbox{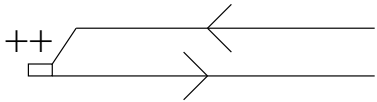}
\end{center}
\caption[]{\small Superposition of the configurations shown in
Fig.\ref{f3}(a) and (d).}
\label{f4}
\end{figure}

Our first task will be to find appropriate conformal field theory
description of these BPS configurations, which 
we shall do in section
\ref{s2}. The construction is best described by using the
boundary state formalism discussed in ref.\cite{BERGAB}
(following earlier work of \cite{POLCAI,CANAYO,MLI}) in which
to each D-brane we associate a coherent state in the closed
string sector describing the wave-function of a closed string emitted
from the brane. (For other applications of this formalism see
\cite{OTHER}. Application of the boundary state formalism to
D-branes moving on orbifolds has been previously discussed in
\cite{IEN}.) The allowed set of D-branes in the theory are
then classified by classifying the possible consistent boundary
states satisfying appropriate symmetry requirements. In our
analysis we find precisely four possible consistent boundary
states, corresponding to the four D-string configurations shown
in Fig.\ref{f3}. The existence of these BPS
configurations provide further
support to the conjecture that the orbifold plane describes the
dual of a D5-brane on top of an O5-plane.

Once we have constructed the boundary state
describing these elementary configurations, we can also
superpose them. In particular, if we superpose the configurations
given in Fig.\ref{f3} (a) and (d), we get a state carrying charge
$2$ under the twisted sector gauge field. This configuration is
shown in Fig.\ref{f4}. The
Ramond-Ramond two form charge, carried by the individual
D-strings, cancel between the two strings. Thus this state has
the same quantum number as the S-dual of the state displayed in
Fig.\ref{f1}(a). However this still has infinite mass from the
tension of the two semi-infinite D-strings and hence cannot be
interpreted as a state living on the orbifold plane. Furthermore
this system has a tachyonic field living on the D-string
world-volume from open strings stretched between the
D-string and the anti-D-string and hence is expected to be
unstable. We interprete this instability as being due to the fact
that the D-string anti D-string system of Fig.\ref{f4} is
unstable against decay into a single ++ state living on the
orbifold plane.
\begin{figure}[!ht]
\begin{center}
\epsfbox{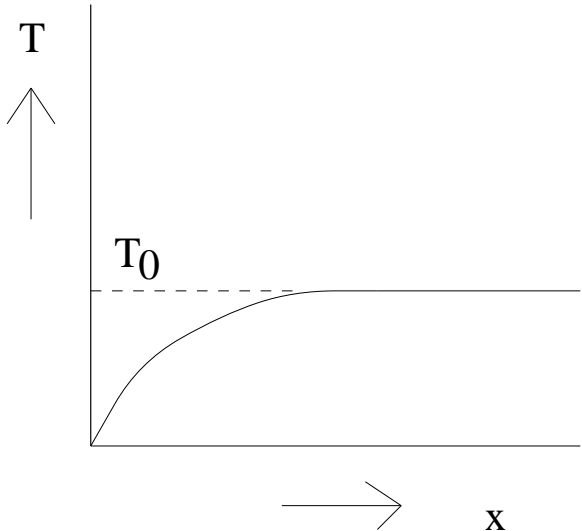}
\end{center}
\caption[]{\small The tachyon field on the D-string - anti-D-string 
pair as a function of distance $x$ from the orbifold
plane.}
\label{f8}
\end{figure}

It is tempting to speculate that the ++ state can actually be
regarded as the state of the D-string anti-D-string pair
displayed in Fig.\ref{f4} after tachyon condensation. If $T_0$ be
the value of the tachyon that minimizes the potential, then far
away from the orbifold plane the tachyon field must approach the
value $T_0$. On the other hand the analysis of section \ref{s23}
tells us that the tachyon field must vanish on the orbifold
plane. Thus the minimum energy configuration will be given by a
tachyon field configuration of the form shown in Fig.\ref{f8}.
If this configuration is to describe the ++ state, then the
negative vacuum energy density on the D-string due to tachyon
condensation far away from the orbifold plane must 
exactly cancel the contribution coming from the D-string
tension\footnote{This in turn implies that space-time
supersymmetry is restored fully in the bulk of the D-string 
anti-D-string pair after
tachyon condensation. Examples where supersymmetry is restored as
a result of tachyon condensation have been discussed previously
in refs.\cite{POLSTR,NARA}.} since otherwise this will give
infinite mass. If this conjecture is correct, then the mass of
the ++ state, measured in string metric, will be of the form
$C/(g\sqrt{\alpha'})$ where $g$ is the string coupling constant,
and $C$ is some numerical constant.
Unfortunately, in the absence of exact knowledge of 
the tachyon
potential we cannot verify the validity of this conjecture, or
proceed further along this line to calculate the numerical
coefficient $C$.\footnote{If instead we consider the superposition of
configurations in Fig.\ref{f3}(a) and \ref{f3}(b), then the
boundary condition on the tachyon near the orbifold plane
requires its derivative along the string to vanish. Thus the
lowest energy configuration corresponds to $T=T_0$ everywhere.
According to our conjecture this will have zero energy. Since
this configuration does not carry any charge and has the same
quantum number as the vacuum, this is what we should
expect.}$^,$\footnote{Non-BPS bound states in quantum field
theory have been discussed recently in \cite{SMILGA}.}
\begin{figure}[!ht]
\begin{center}
\epsfbox{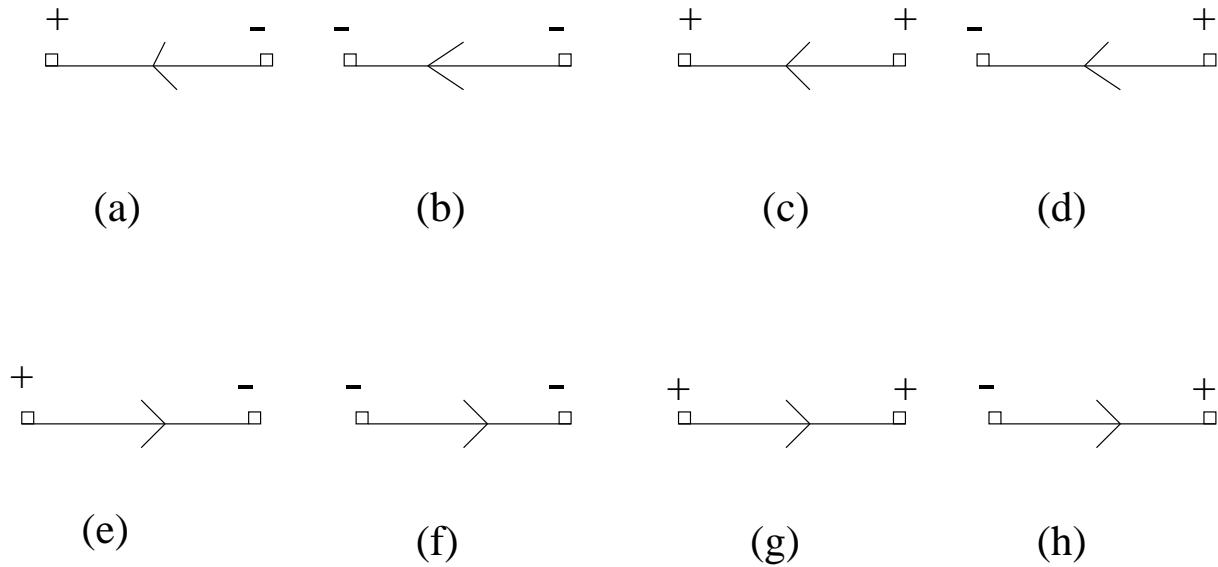}
\end{center}
\caption[]{\small The eight different configurations of D-strings
stretched between the two orbifold planes in type IIB on
$R^{8,1}\times S^1/\II_4\cdot (-1)^{F_L}$.}
\label{f5}
\end{figure}

In order to gain further insight into this problem, we consider in
section \ref{s3} type IIB string theory on an orbifold
$R^{8,1}\times S^1/(-1)^{F_L}\cdot \II_4$ where $\II_4$ reverses
the sign of the coordinate on $S^1$ and three other space-like
directions. This orbifold has two fixed planes, separated by a
distance $\pi R$ where $R$ is the radius of the circle. As shown
in Fig.\ref{f5}, this
system is expected to have eight different BPS configurations 
corresponding to D-strings stretched between the two orbifold planes.
We explicitly construct these D-brane configurations by finding the
appropriate boundary state in the closed string sector for each
of these brane configurations. Each of these configurations has
mass $\pi R T_D$ where $T_D$ is the D-string tension.
\begin{figure}[!ht]
\begin{center}
\epsfbox{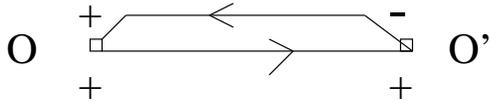}
\end{center}
\caption[]{\small Superposition of the D-brane configurations of
Fig.\ref{f5}(a) and (g).}
\label{f6}
\end{figure}

We can also find the boundary
states corresponding to the superposition of any number of these
D-branes by superposing the corresponding boundary states. Let us
consider the superposition of the configurations in
Fig.\ref{f5}(a) and \ref{f5}(g). The corresponding state is
shown in Fig.\ref{f6} and carries the same quantum numbers as the
state ++ living on the orbifold plane $O$. Since this
configuration has total mass $2\pi R T_D$, we
expect that for large $R$ this configuration will
be unstable under decay into the
state ++ living on the orbifold plane, as the latter has a finite
mass independent of $R$. 
On the other hand, for small $R$ the
total mass of the system shown in Fig.\ref{f6} goes to zero and
we expect the system to be stable. Thus if our interpretation of
the tachyonic instability is correct, then for large $R$ the
system should have tachyonic modes, but for sufficiently small
$R$ there should be no tachyonic mode. This is indeed shown to be
the case by analyzing the spectrum of open strings for this
system.

Thus we arrive at the following scenario. For sufficiently small
radius, the system shown in Fig.\ref{f6} is stable and describes
the lowest mass configuration with these quantum numbers. As we
increase $R$ beyond a critical radius $R_c$, a tachyonic mode
appears and the system becomes unstable due to the possibility of
decaying into the ++ state.\footnote{In fact at $R=R_c$ the
`tachyon' becomes an exactly marginal operator in the underlying
conformal field theory.} Thus the mass of the D-string anti-
D-string system at $R=R_c$ must be equal to the mass $m_{++}$ of
the ++ state living on the orbifold plane. This gives
\be \label{ei1}
m_{++} = 2 \pi R_c T_D\, ,
\ee
where $T_D$ is the tension of the D-string. Measured in the
string metric, $T_D=(2\pi\alpha'g)^{-1}$ where $g$ is the 
string coupling constant. Explicit calculation
shows that  in this case $R_c = \sqrt{\alpha'/2}$.  This gives
\be \label{ei2}
m_{++} = (\sqrt{2\alpha'}g)^{-1} .
\ee
\begin{figure}[!ht]
\begin{center}
\epsfbox{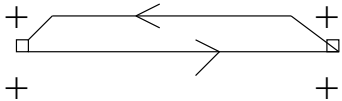}
\end{center}
\caption[]{\small Superposition of the D-brane configurations of
Fig.\ref{f5}(c) and (g).}
\label{f7}
\end{figure}

This result should be interpreted with care, 
since $m_{++}$ may be affected by the presence of
the second orbifold plane at a distance $\pi R_c$ away, and hence
the mass $m_{++}$ calculated this way need not be the same as the
mass $m_{++}$ of a doubly charged state living on an isolated
orbifold plane.  Although
we do not have a direct knowledge of the $R$ dependence of $m_{++}$,
we have indirect evidence that in the $g\to 0$ limit $m_{++}$
might indeed be independent of $R$. This is seen by considering
the superposition of the D-string configurations shown in
Figs.\ref{f5}(c) and \ref{f5}(g), as shown in Fig.\ref{f7}. Again we find
that for small $R$ there is no tachyon, whereas for large $R$
there is a tachyon signalling the decay of this state into a ++
state living on the first orbifold plane, and a ++ state living
on the second orbifold plane. The critical radius where the
tachyon appears turns out to be $R_c'=\sqrt{2\alpha'}$, and hence
the total mass of the pair of D-strings at this critical radius
is given by $2\pi R_c' T_D=(g\sqrt{\alpha'/2})^{-1}$. This should
be equal to $2m_{++}$.\footnote{Since there are a pair of ++
states, one living on each orbifold, there will also be a
contribution to the mass
due to the interaction between these two states.  However,
since each state has mass of order $g^{-1}$ and the Newton's
constant is of order $g^2$, this interaction energy will be of
order unity, and will be negligible compared to the total
mass of the system in the $g\to 0$ limit.}
This gives the same value of $m_{++}$ as
given in eq.\refb{ei2} although now the separation between the two
orbifold planes is double of what it was in the previous case.
Since $m_{++}$ calculated this way has the same value for $R=R_c$
and $R=R_c'$, we have reason to believe that it might be
independent of $R$ and gives the value of $m_{++}$ for $R\to
\infty$.

Given the answer \refb{ei2} for $m_{++}$ one can reexpress it in
terms of the variables of the dual theory. If $\wt g$ is the
coupling constant of the dual theory, then the mass of this state,
measured in the string metric of the dual theory, turns out to
be: 
\be \label{ei3}
\wt m_{++} = (\sqrt{2\alpha'})^{-1} \wt g^{1/2}.
\ee
This will be the mass of a ++ state living on the coincident
D5-brane $-$ O5-plane system in the limit of large $\wt g$.

\sectiono{D-String Ending on an Orbifold Plane} \label{s2}

In this section we shall construct the boundary states describing
D-strings ending on an orbifold plane for each of the four
configurations displayed in Fig.\ref{f3}, as well as their
superposition shown in Fig.\ref{f4}. We begin in subsection
\ref{s21} by reviewing
the construction of the boundary state describing a D-string
in type IIB string theory in flat ten dimensional Minkowski
space. In subsection \ref{s22} we extend this construction to a
D-string ending on the orbifold plane. In subsection \ref{s23} we
analyze the dynamics of a configuration obtained by superposition
of two such D-string configurations.

\subsection{Review of D-strings in Flat Space} \label{s21}

Let us consider a configuration of D-string in type IIB string
theory on $R^{9,1}$. We shall denote the coordinates of $R^{9,1}$
by $x^0, \ldots x^9$, and take the D-string to be stretched along
$x^9$ and situated at $x^1=\ldots x^8=0$. 
A convenient description of the D-string is given in terms
of boundary states describing the wave-function of a closed string 
emitted from the D-string\cite{POLCAI,CANAYO,MLI,BERGAB}. We work
in the notation of \cite{BERGAB} by making a double wick rotation
$x^0\to ix^0$, $x^1\to ix^1$, so that the directions tangential
to the D-string world-sheet are space-like, and then go to the
light-cone gauge by taking $x^1\pm x^2$ as the light-cone
directions.\footnote{The advantage of the light-cone gauge
is that we avoid having to deal with the ghost sector. However
since the treatment of this sector for D-branes and the usual
string theory is identical, we could easily work 
in the fully covariant formulation 
following \cite{POLCAI,CANAYO,MLI}.}
We now define the coherent states:
\ben \label{e1}
|k,\eta\ra_{NSNS\atop RR} 
&=& \exp\Big( \sum_{n=1}^\infty \big[
-{1\over n}\sum_{\mu = 0,9}\alpha^\mu_{-n} \wt \alpha^\mu_{-n}
+{1\over n}\sum_{\mu =3}^8\alpha^\mu_{-n} \wt
\alpha^\mu_{-n}\big] \nonumber \\
&& + i\eta \sum_{r>0} \big[ 
-\sum_{\mu=0,9} \psi^\mu_{-r}\wt\psi^\mu_{-r}
+\sum_{\mu=3}^8 \psi^\mu_{-r}\wt\psi^\mu_{-r}\big]\Big)
|k,\eta\ra_{NSNS\atop RR}^{(0)}\, ,
\een
where $k$ denotes the eight dimensional momentum $(k^1,\ldots
k^8)$ transverse to the D-string, $\alpha^\mu_n$ and
$\wt\alpha^\mu_n$ are the right- and the left-moving modes of the
bosonic coordinate $X^\mu$, $\psi^\mu_r$ and $\wt\psi^\mu_r$
are the modes of the right- and the left-moving fermionic
coordinates $\psi^\mu$, $\wt\psi^\mu$,
and $\eta$ can take values $\pm 1$. Our notations and
normalization conventions are described in appendix \ref{sa};
also we have set the string tension to $1/2\pi$ ($\alpha'=1$). 
The sum over $n$
runs over positive integers, whereas the sum over $r$ runs over
positive integers (integers + ${1\over 2}$) in the RR (NSNS)
sector. 
$|k,\eta\ra^{(0)}$ denotes the Fock vacuum carrying momentum
$k$ in directions transverse to the D-string and zero momentum
along directions tangential to the D-string.
In the NSNS sector this uniquely specifies this state and hence
it is independent of $\eta$. The situation in the RR sector is a
bit more complicated due to the presence of the fermionic zero
modes. In order to
define $|k,\eta\ra^{(0)}_{RR}$ we define:
\be \label{e2}
\psi^\mu_\pm = {1\over \sqrt 2} (\psi^\mu_0\pm i \wt
\psi^\mu_0)\, , \qquad \mu = 0, 3,4\ldots 9\, ,
\ee
where $\psi_0^\mu$ and $\wt\psi_0^\mu$ are the zero modes of the
left- and the right-moving modes of the world-sheet fermions,
normalized as in appendix \ref{sa}. We now define 
$|k,-\ra_{RR}^{(0)}$ to be the RR ground state satisfying:
\ben \label{e3}
\psi^\mu_-|k,-\ra^{(0)}_{RR} &=& 0 \qquad \hbox{for} \quad
\mu=0,9\nonumber \\
\psi^\mu_+|k,-\ra^{(0)}_{RR} &=& 0 \qquad \hbox{for} \quad
\mu=3,4\ldots 8\, .
\een
$|k,+\ra^{(0)}_{RR}$ is now defined as
\be \label{e4}
|k,+\ra^{(0)}_{RR}=\prod_{\mu=3}^8\psi^\mu_- \prod_{\mu=0,9}\psi^\mu_+
\, |k,-\ra^{(0)}_{RR}\, .
\ee

Let us now define:
\ben \label{e5}
|\eta\ra_{NSNS} &=& \NN \int \Big(\prod_{\mu=1}^8 dk^\mu\Big) \,
|k,\eta\ra_{NSNS}\, , \nonumber \\
|\eta\ra_{RR} &=& 4i\NN \int \Big(\prod_{\mu=1}^8 dk^\mu\Big) \,
|k,\eta\ra_{RR}\, , 
\een
\ben \label{e6}
|U\ra_{NSNS} & = & {1\over \sqrt 2}
(|+\ra_{NSNS} - |-\ra_{NSNS})\, , \nonumber \\
|U\ra_{RR} & = & {1\over\sqrt 2}(|+\ra_{RR} + |-\ra_{RR})\, , 
\een
\ben \label{e7}
|D1\ra & = & {1\over \sqrt 2}(|U\ra_{NSNS} + |U\ra_{RR})
\, , \nonumber \\
|\wb{D1}\ra & = & {1\over \sqrt 2} (|U\ra_{NSNS} - 
|U\ra_{RR})\, .
\een
In eqs.\refb{e6} and \refb{e7} the symbol $U$ stands for
untwisted sector; although at present we are dealing with
D-strings in flat space, we have attached these symbols in
anticipation of the analysis in the next subsection. $\NN$ is a
normalization constant which will be fixed later. The states
$|\eta\ra_{NSNS}$ and $|\eta\ra_{RR}$ defined in eq.\refb{e5}
can be shown to satisfy the
boundary conditions relevant for a D-string situated at $x^\mu=0$
($1\le \mu \le 8$):
\ben \label{e8}
X^\mu(\tau=0,\sigma) 
|\eta\ra_{NSNS\atop RR} &=& 0 \qquad \hbox{for}
\quad 3\le \mu\le 8\, ,\nonumber \\
\p_\tau
X^\mu(\tau=0,\sigma) |\eta\ra_{NSNS\atop RR} &=& 0 \qquad \hbox{for}
\quad \mu = 0,9 \, , \nonumber \\
(\psi^\mu - i\eta \wt\psi^\mu)    
|\eta\ra_{NSNS\atop RR} &=& 0 \qquad \hbox{for}
\quad 3\le \mu\le 8\, ,\nonumber \\
(\psi^\mu +i\eta\wt\psi^\mu)
|\eta\ra_{NSNS\atop RR} &=& 0 \qquad \hbox{for}
\quad \mu = 0,9 \, , \nonumber \\
x^\mu |\eta\ra_{NSNS\atop RR} &=& 0 \qquad \hbox{for}
\quad \mu = 1,2 \, . 
\een
The states $|U\ra_{NSNS}$ and $|U\ra_{RR}$ defined in
eq.\refb{e6} satisfy GSO projection, {\it i.e.} they are
invariant under the operators $(-1)^F$ and $(-1)^{\wt F}$ defined
in appendix \ref{sa}.   Finally, the states
$|D1\ra$ and $|\wb{D1}\ra$ defined in eq.\refb{e7} denote 
the boundary
states associated with a D-string and an anti-D-string
respectively. (An anti-D-string can be identified with a
D-string with opposite orientation.) Note that although the
states $|U\ra_{NSNS}$ and $|U\ra_{RR}$ by themselves satisfy the
GSO projection as well as the boundary condition \refb{e8}, we
need to take the specific linear combinations defined in
eq.\refb{e7} for satisfying the open-closed consistency
condition\cite{BERGAB} that the amplitude for emission and
reabsorption of a closed string from the brane must be
interpretable as the partition function of open string states
satisfying appropriate GSO projection. We shall now see
explicitly how this happens.

The
amplitude for the emission and reabsorption of a closed string
from the D-string is given by
\be \label{e9}
\int_0^\infty dl \,  \la D1|e^{-lH_c}|D1\ra \, ,
\ee
where $H_c$ is the hamiltonian for closed strings in the light-cone
gauge:
\be \label{e10}
H_c = \pi p^2 + 2\pi \sum_{\mu=0,3,\ldots 9}
\Big[\sum_{n=1}^\infty
(\alpha^\mu_{-n}\alpha^\mu_n +\wt\alpha^\mu_{-n}\wt\alpha^\mu_n)
+ \sum_{r>0} r (\psi^\mu_{-r}\psi^\mu_r
+\wt\psi^\mu_{-r}\wt\psi^\mu_r)\Big] + 
2\pi C_c\, .
\ee
The constant $C_c$ takes the value $-1$ in the NSNS sector, and zero in
the RR sector. \refb{e9} can be evaluated with the help
of the following identities:\footnote{The factor of $l^{-4}$ in these 
equations comes from momentum integration.}$^,$\footnote{In 
defining the bra vectors we use the convention that the factor of $i$
appearing in eq.\refb{e5} is not conjugated.}
\ben \label{e11}
&& {}~_{NSNS}\la + | e^{-lH_c}| +\ra_{NSNS}
= {}~_{NSNS}\la - | e^{-lH_c}|-\ra_{NSNS} =
\NN^2 l^{-4} {f_3(q)^8\over
f_1(q)^8}\, , \nonumber \\
&& {}~_{NSNS}\la + | e^{-lH_c}| -\ra_{NSNS}
= {}~_{NSNS}\la - |e^{-lH_c}| +\ra_{NSNS} = 
\NN^2 l^{-4} {f_4(q)^8\over
f_1(q)^8}\, , \nonumber \\
&& {}~_{RR}\la + |e^{-lH_c}| +\ra_{RR}
= {}~_{RR}\la - | e^{-lH_c}|  -\ra_{RR} = - 
\NN^2 l^{-4} {f_2(q)^8\over
f_1(q)^8}\, , \nonumber \\
&& {}~_{RR}\la + | e^{-lH_c}| -\ra_{RR}
= {}~_{RR}\la - | e^{-lH_c}| +\ra_{RR} = 0\, ,
\een
where $q\equiv \exp(-2\pi l)$ and\cite{POLCAI},
\ben \label{e12}
f_1(q) &=& q^{1/12} \prod_{n=1}^\infty (1-q^{2n})\, , \nonumber
\\
f_2(q) &=& \sqrt 2\, q^{1/12} \prod_{n=1}^\infty (1+q^{2n})\, , \nonumber
\\
f_3(q) &=& q^{-1/24} \prod_{n=1}^\infty (1+q^{2n-1})\, , \nonumber
\\
f_4(q) &=& q^{-1/24} \prod_{n=1}^\infty (1-q^{2n-1})\, .
\een
The functions $f_i$ have the following modular transformation
properties:
\ben \label{e13}
&& f_1(e^{-\pi/t})=\sqrt t f_1(e^{-\pi t})\, , \qquad 
f_2(e^{-\pi/t})= f_4(e^{-\pi t})\, , \nonumber \\ 
&& f_3(e^{-\pi/t})= f_3(e^{-\pi t})\, , \qquad 
f_4(e^{-\pi/t})= f_2(e^{-\pi t})\, .
\een
Let $H_o$ be the Hamiltonian for an open string with ends lying on the
D-brane
\be \label{e15}
H_{o} = \pi p^2 + \pi \sum_{\mu=0,3,\ldots 9}
\Big[\sum_{n=1}^\infty
\alpha^\mu_{-n}\alpha^\mu_n 
+ \sum_{r>0} r \psi^\mu_{-r}\psi^\mu_r
\Big] + \pi C_{o}\, .
\ee
$C_{o}$ vanishes in the Ramond sector and takes the value
$-1/2$ in the NS sector. $p$ denotes the open string momentum tangential
to the D-string. Let us denote by $Tr_{NS}$ the trace over the 
NS sector states of the open string Hilbert space and
$Tr_R$ the trace over the Ramond sector states of
the open string Hilbert space. These traces include an
integration over the momentum tangential to the D-string,
weighted by the density of states $A/4\pi^2$, where $A$ is the
(infinite) area of the $x^0-x^9$ plane. {\it Also both traces are
taken without the GSO projection.}  Defining new  variables
$t=(2l)^{-1}$, $\wt q=e^{-\pi t}$, and using 
\refb{e11}-\refb{e15} we get
\ben \label{e14a}
&& \int_0^\infty dl \,  ~_{NSNS}\la + | e^{-lH_c}| +\ra_{NSNS}
= \int_0^\infty dl \,  {}~_{NSNS}\la - | e^{-lH_c}|-\ra_{NSNS}
\nonumber \\ & = &
8\NN^2 \int_0^\infty {dt\over t^2} \Big({f_3(\wt q)\over f_1(\wt
q)}\Big)^8 
=  \int_0^\infty {dt\over 2t} Tr_{NS} (e^{-2tH_{o}})
\, , 
\een
\ben \label{e14b}
&& \int_0^\infty dl \,  ~_{NSNS}\la + | e^{-lH_c}| -\ra_{NSNS}
= \int_0^\infty dl \,  {}~_{NSNS}\la - | e^{-lH_c}|+\ra_{NSNS}
\nonumber \\ & = &
8 \NN^2 \int_0^\infty {dt\over t^2} \Big({f_2(\wt q)\over f_1(\wt
q)}\Big)^8 
=  \int_0^\infty {dt\over 2t} Tr_{R}(e^{-2tH_{o}})
\, , 
\een
\ben \label{e14c}
&& \int_0^\infty dl \,  ~_{RR}\la + | e^{-lH_c}| +\ra_{RR}
= \int_0^\infty dl \,  {}~_{RR}\la - | e^{-lH_c}|-\ra_{RR} \nonumber \\
& = &
- 8\NN^2 \int_0^\infty {dt\over t^2} \Big({f_4(\wt q)\over f_1(\wt
q)}\Big)^8 
= \int_0^\infty {dt\over 2t} Tr_{NS}(e^{-2tH_{o}}
(-1)^F)
\, , 
\een
\be \label{e14d}
\int_0^\infty dl \,  ~_{RR}\la + | e^{-lH_c}| -\ra_{RR}
= \int_0^\infty dl \,  {}~_{RR}\la - | e^{-lH_c}|+\ra_{RR} = 0 \, ,
\ee
provided we choose
\be \label{e16}
32 \NN^2 = {A\over 4\pi^2}\, .
\ee
The $-$ sign in eq.\refb{e14c} shows that the open
string NS sector ground state is odd under $(-1)^F$.

Using eqs.\refb{e6}, \refb{e7},
and \refb{e14a}-\refb{e14d} we can now evaluate \refb{e9}:
\be \label{ep1}
\int_0^\infty dl \,  \la D1| e^{-lH_c}| D1\ra = \int_0^\infty 
{dt\over 2t} \, \Big[
Tr_{NS} \Big( e^{-2t H_o} {1 + (-1)^F\over 2}\Big) -
Tr_{R} \Big( e^{-2t H_o} {1 + (-1)^F\over 2}\Big)\Big]\, ,
\ee
where we have used the identity:
\be \label{ep1a}
Tr_R\Big( e^{-2tH_o} (-1)^F\Big) =0\, .
\ee
\refb{ep1} can be identified as the one loop partition function
of open strings with both ends lying on the D-string.

Let us now consider a D-string anti-D-string pair situated at
$x^1=\ldots x^8=0$. This system is described by a boundary state:
\be \label{ep2}
|D1;\wb{D1}\ra \equiv |D1\ra + |\wb{D1}\ra\, .
\ee
The corresponding amplitude for the emission and reabsorption of a
closed string from this D-string pair is given by
\be \label{ep3}
\int_0^\infty dl \,  \la D1;\wb{D1}| e^{-lH_c}|D1;\wb{D1}\ra\, .
\ee
This can be easily evaluated using eqs.\refb{e14a}-\refb{e14d} by
noting that $|D1\ra + |\wb{D1}\ra=\sqrt 2 |U\ra_{NSNS}$. However
in order to give a physical interpretation of the result it is
more convenient to rewrite this integral as:
\be \label{ep4}
\int_0^\infty dl \,  \big[ \la D1 | e^{-lH_c}| D1\ra 
+ \la \wb{D1} | e^{-lH_c}| \wb{D1}\ra 
+ \la {D1} | e^{-lH_c}| \wb{D1}\ra 
+ \la \wb{D1} | e^{-lH_c}| {D1}\ra \big]\, .
\ee
Each of the first two terms are given by \refb{ep1} and can be
interpreted as the partition function of open strings with both
ends lying on the D-string and both ends lying on the 
anti-D-string respectively. On the other hand, using
eqs.\refb{e6}, \refb{e7}, \refb{e14a}-\refb{e14d}, \refb{ep1a} 
we get
\be \label{ep5}
\int_0^\infty dl \,  \la D1| e^{-lH_c}| \wb{D1}\ra =
\int_0^\infty dl \,  \la \wb{D1}| e^{-lH_c}| D1\ra =
\int_0^\infty {dt\over 2t}
Tr_{NS-R} \Big( e^{-2t H_o} {1 - (-1)^F\over 2}\Big) \, ,
\ee
where $Tr_{NS-R}$ denotes the difference between the traces in
the NS and the R sector of open strings.
These have to be interpreted as the partition functions of open
strings starting on the D-string and ending on the anti-D-string
and vice versa. The $-$ sign in front of the $(-1)^F$ term in
the partition function shows that these open strings have opposite GSO
projection compared to the open strings with both ends ending on
D-string or anti-D-string. Since the NS sector ground state has
$(-1)^F=-1$, we see that it now survives the GSO projection,
giving rise to the tachyonic mode living on the D-string - 
anti-D-string world-sheet\cite{BANSUS,GRGUP,PERI}. 

We can formalise this by assigning
$2\times 2$ Chan-Paton matrices to the open strings
living on the world volume of the D-string anti-D-string pair.
The open strings with both ends lying on the D-string or 
anti-D-string correspond
to diagonal Chan-Paton matrices $\pmatrix{1 & 0\cr 0 & 0}$ and
$\pmatrix{0 & 0\cr 0 & 1}$, whereas the open strings with two
ends lying on the two different branes correspond to off-diagonal
Chan-Paton matrices $\pmatrix{0 & 1\cr 0 & 0}$ and $\pmatrix{0 &
0\cr 1 & 0}$. The action of $(-1)^F$ on the Chan-Paton factor is
taken to be
conjugation by the matrix $\pmatrix{1 & \cr & -1}$. With this
convention, the total wave-function of any physical
open string state is 
always invariant under $(-1)^F$.

\subsection{D-string on a $Z_2$ Orbifold} \label{s22}

Let us now consider a $Z_2$ orbifold of the D-string
configuration discussed in the previous section. The orbifold
group is generated by the element:
\be \label{ep6}
g = (-1)^{F_L}\cdot \II_4\, ,
\ee
where $\II_4$ changes the sign of the coordinates $x^6,\ldots
x^9$, and $(-1)^{F_L}$ changes the sign of the Ramond sector
ground state on the left without having any action on any of the
oscillators, or the ground state of any other sector. Note that
the D-string along the 9th direction is transverse to the
orbifold fixed plane spanned by $x^1,\ldots x^5$.  Our first
task will be to verify that the states $|U\ra_{NSNS}$ and
$|U\ra_{RR}$ are invariant under $g$. 
In both, the NSNS and the RR sectors,
$\II_4$ induces the transformation
\be \label{ep7}
\alpha^\mu_n\to -\alpha^\mu_n, \quad 
\wt\alpha^\mu_n\to -\wt\alpha^\mu_n, \quad 
\psi^\mu_r\to -\psi^\mu_r, \quad 
\wt\psi^\mu_r\to -\wt\psi^\mu_r,  \qquad
\hbox{for} \quad 6\le \mu\le 9 \, .
\ee
$\II_4$ acts on the Fock vacuum in the NSNS sector as
\be \label{ep7a}
|k^0, \ldots k^9\ra^{(0)}_{NSNS} \to |k^0, \ldots k^5, -k^6, 
\ldots -k^9\ra^{(0)}_{NSNS}\, .
\ee
On the other hand $(-1)^{F_L}$ has no action in the NSNS sector.
Thus $|\eta\ra_{NSNS}$ defined in  eqs. \refb{e1},
\refb{e5} and hence $|U\ra_{NSNS}$ defined in eq.\refb{e6} is
invariant under $g$.

The action of $\II_4$ on the RR sector ground state takes the
form:
\be \label{ep8}
|k^0, \ldots k^9, \eta\ra^{(0)}_{RR}
\to \prod_{\mu=6}^9 (\sqrt 2 \psi_0^\mu)
\prod_{\mu=6}^9 (\sqrt 2 \wt\psi_0^\mu)
|k^0, \ldots k^5, -k^6, \ldots -k^9, \eta\ra^{(0)}_{RR}\, .
\ee
On the other hand $(-1)^{F_L}$ acting on the RR ground state
changes its sign. Using these relations one can verify that
$|U\ra_{RR}$ defined in eq.\refb{e6}
is also invariant under $g$.

Since both $|U\ra_{NSNS}$ and $|U\ra_{RR}$ are invariant under
$g$, the boundary states $|D1\ra$ and $|\wb{D1}\ra$ defined in
\refb{e7} are also invariant under $g$. However, this does not
mean that the boundary state of a D-string (anti-D-string)
stretched along the 9th direction is given by $|D1\ra$
($|\wb{D1}\ra$). To see the reason for this, we note that 
the state $|D1\ra$ satisfies eq.\refb{ep1}, but the correct
boundary state $|D1'\ra$ should give an answer where inside
the trace we have a projection operator projecting onto the
subspace of $g$ invariant open string states:
\be \label{ep10}
\int_0^\infty dl \,  \la D1'| e^{-lH_c}| D1'\ra 
= \int_0^\infty {dt\over 2t}
Tr_{NS-R} \Big( e^{-2t H_o} {1 + (-1)^F\over 2}{1+g\over 2}\Big)
\, .
\ee
As we shall now show, the remedy lies in modifying the
definitions of $|D1\ra$, $|\wb{D1}\ra$ by adding to them
appropriate coherent states from the twisted sector satisfying
equations analogous to \refb{e8}. 
For this we define the analog
of \refb{e1} in the twisted sector:
\ben \label{ep11}
|k,\eta\ra_{NSNS;T\atop RR;T} 
&=& \exp\Big( \sum_{n>0} \big[
-{1\over n}\sum_{\mu = 0,9}\alpha^\mu_{-n} \wt \alpha^\mu_{-n}
+{1\over n}\sum_{\mu =3}^8\alpha^\mu_{-n} \wt
\alpha^\mu_{-n}\big] \nonumber \\
&& + i\eta \sum_{r>0} \big[ 
-\sum_{\mu=0,9} \psi^\mu_{-r}\wt\psi^\mu_{-r}
+\sum_{\mu=3}^8 \psi^\mu_{-r}\wt\psi^\mu_{-r}\big]\Big)
|k,\eta\ra_{NSNS;T\atop RR;T}^{(0)}\, ,
\een
where $k\equiv (k^1,\ldots k^5)$ denotes the five dimensional
momentum transverse to the D-string and tangential to the
orbifold plane. In the NSNS sector the integers $n$ 
and $r$ take values
\ben \label{ep12}
n &\in& Z_+ \qquad \hbox{for} \qquad \mu=0,3,4,5 \nonumber \\
&\in& Z_+ - {1\over 2} \qquad \hbox{for} 
\qquad \mu=6,7,8,9 \nonumber \\
r &\in& Z_+ -{1\over 2} \qquad \hbox{for} 
\qquad \mu=0,3,4,5 \nonumber \\
&\in& Z_+ \qquad \hbox{for} 
\qquad \mu=6,7,8,9 \, .
\een
On the other hand, in the RR sector 
\ben \label{ep13}
n,r &\in& Z_+ \qquad \hbox{for} \qquad \mu=0,3,4,5 \nonumber \\
&\in& Z_+ - {1\over 2} \qquad \hbox{for} 
\qquad \mu=6,7,8,9 \, .
\een
Here $Z_+$ denotes the set of positive integers.

Note that there are fermion zero modes in both the NSNS and the
RR sector. Hence we must carefully define the states 
$|k,\eta\ra_{NSNS;T\atop RR;T}^{(0)}$ in both sectors. In the
NSNS sector the fermion zero modes are $\psi_0^\mu$ and
$\wt\psi_0^\mu$ for $6\le\mu\le 9$. We define $\psi_\pm^\mu$ as
in eq.\refb{e2} for these values of $\mu$, and the ground
state $|k,-\ra^{(0)}_{NSNS;T}$ satisfies:
\ben \label{ep14}
\psi^\mu_-|k,-\ra^{(0)}_{NSNS;T} &=& 0 \qquad \hbox{for} \quad
\mu=9\nonumber \\
\psi^\mu_+|k,-\ra^{(0)}_{NSNS;T} &=& 0 \qquad \hbox{for} \quad
\mu=6,7,8\, .
\een
$|k,+\ra^{(0)}_{NSNS;T}$ is now defined as
\be \label{ep15}
|k,+\ra^{(0)}_{NSNS;T}=\Big(\prod_{\mu=6}^8\psi^\mu_-\Big)\,
\psi^9_+ |k,-\ra^{(0)}_{NSNS;T}\, .
\ee
In the RR sector the fermion zero modes are $\psi_0^\mu$,
$\wt\psi_0^\mu$ for $\mu=0,3,4,5$. We again define $\psi^\mu_\pm$
for these values of $\mu$ using eq.\refb{e2} and the ground
state $|k,-\ra^{(0)}_{RR;T}$ satisfies: 
\ben \label{ep16}
\psi^\mu_-|k,-\ra^{(0)}_{RR;T} &=& 0 \qquad \hbox{for} \quad
\mu=0\nonumber \\
\psi^\mu_+|k,-\ra^{(0)}_{RR;T} &=& 0 \qquad \hbox{for} \quad
\mu=3,4,5\, .
\een
$|k,+\ra^{(0)}_{RR;T}$ is defined as
\be \label{ep17}
|k,+\ra^{(0)}_{RR;T}=\Big(\prod_{\mu=3}^5 
\psi^\mu_-\Big)\, \psi^0_+ |k,-\ra^{(0)}_{RR;T}\, .
\ee
We now define
\be \label{ep18}
|\eta\ra_{NSNS;T} = 2\wt \NN \int \Big(\prod_{\mu=1}^5
dk^\mu\Big) |k,\eta\ra_{NSNS;T}\, ,
\ee
\be \label{ep19}
|\eta\ra_{RR;T} = 2i\wt \NN \int \Big(\prod_{\mu=1}^5
dk^\mu\Big) |k,\eta\ra_{RR;T}\, ,
\ee
where $\wt \NN$ is a suitable normalisation constant to be
determined later. 

The action of the GSO operators $(-1)^F$ and $(-1)^{\wt F}$ in
the ground state of the NSNS sector are given by:
\be \label{ep20}
(-1)^F: \quad \prod_{\mu=6}^9 (\sqrt 2\psi_0^\mu), \qquad 
(-1)^{\wt F}: \quad - \prod_{\mu=6}^9 (\sqrt 2 \wt\psi_0^\mu)\, .
\ee
On the other hand, acting on the ground state of the RR sector
these operators take the form:
\be \label{ep21}
(-1)^F: \quad \Big(\prod_{\mu=3}^5 (\sqrt 2 \psi_0^\mu)\Big)
(\sqrt 2\psi_0^0), 
\qquad 
(-1)^{\wt F}: \quad - \Big(\prod_{\mu=3}^5 
(\sqrt 2\wt\psi_0^\mu)\Big)
(\sqrt 2\wt\psi_0^0)\, .
\ee
The overall sign in the definition of each operator is determined
by requiring consistency with the residual space-time
supersymmetry of the orbifold theory. With this choice of sign, the
massless states coming from the NSNS sector transform as a vector
under the `internal' R-symmetry group SO(4) acting on $x^6,
\ldots x^9$, whereas the massless states coming from the RR
sector transform as a vector under the rotation group SO(4)
acting on $x^0,x^3,x^4,x^5$. This is precisely the bosonic part
of the spectrum of a massless vector multiplet living on the
orbifold plane.  Using
eqs.\refb{ep11}-\refb{ep21} we can verify that the following
combinations of the boundary states are invariant under the GSO
projection:
\be \label{ep22}
|T\ra_{NSNS} = {1\over \sqrt 2} (|+\ra_{NSNS;T} +
|-\ra_{NSNS;T})\, ,
\ee
\be \label{ep23}
|T\ra_{RR} = {1\over \sqrt 2} (|+\ra_{RR;T} +
|-\ra_{RR;T})\, ,
\ee

The closed string Hamiltonian in the twisted sector is
given by eq.\refb{e10} with $C_c=0$ in both NSNS and RR sector.
The analog of eqs.\refb{e14a}-\refb{e14d} are now given by,
\ben \label{eq1}
&& \int_0^\infty dl \,  ~_{NSNS;T}\la + | e^{-lH_c}| +\ra_{NSNS;T}
= \int_0^\infty dl \,  {}~_{NSNS;T}\la - | e^{-lH_c}|-\ra_{NSNS;T}
\nonumber \\ & = &
2^{3/2}\wt\NN^2 \int_0^\infty {dt\over t^{3/2}} \Big({f_4(\wt q)
f_3(\wt q)\over f_1(\wt q) f_2(\wt q)}\Big)^4 
=  \int_0^\infty {dt\over 2t} Tr_{NS} ( e^{-2tH_{o}} \cdot g)
\, , 
\een
\ben \label{eq2}
&& \int_0^\infty dl \,  ~_{NSNS;T}\la + | e^{-lH_c}| -\ra_{NSNS;T}
= \int_0^\infty dl \,  {}~_{NSNS;T}\la - | e^{-lH_c}|+\ra_{NSNS;T}
\nonumber \\ & = & 0
= - \int_0^\infty {dt\over 2t} Tr_{R}(e^{-2tH_{o}}\cdot g)
\, , 
\een
\ben \label{eq3}
&& \int_0^\infty dl \,  ~_{RR;T}\la + | e^{-lH_c}| +\ra_{RR;T}
= \int_0^\infty dl \,  {}~_{RR;T}\la - | e^{-lH_c}|-\ra_{RR;T} 
\nonumber \\
& = &
- 2^{3/2} \wt \NN^2 \int_0^\infty {dt\over t^{3/2}} 
\Big({f_3(\wt q) f_4(\wt q)\over f_1(\wt
q) f_2(\wt q)}\Big)^4 
= \int_0^\infty {dt\over 2t} Tr_{NS}(e^{-2tH_{o}} \cdot
g\cdot (-1)^F)
\, , 
\een
\ben \label{eq4}
&& \int_0^\infty dl \,  ~_{RR;T}\la + | e^{-lH_c}| -\ra_{RR;T}
= \int_0^\infty dl \,  {}~_{RR;T}\la - | e^{-lH_c}|+\ra_{RR;T}
\nonumber \\
& = & 0 
= - \int_0^\infty {dt\over 2t} Tr_{R}(e^{-2tH_{o}} \cdot
g\cdot (-1)^F) \, .
\een
provided we choose
\be \label{eq4a}
2 \wt \NN^2 = {L\over 2\pi}\, ,
\ee
where $L$ is the (infinite) length of the $x^0$ direction.
Conjugation by $g$ changes the sign of the open string
oscillators associated with the 6,7,8,9 directions.

There are some subtle points in eqs.\refb{eq1}-\refb{eq4}
which we discuss below:
\begin{enumerate}
\item Although eqs.\refb{eq1} and \refb{eq3} have identical
expressions up to a sign, various terms in these 
expressions have different
origin in the two cases. For example, in \refb{eq1}
the factor of $(f_4(\wt q))^4$ in the numerator comes from the
fermions associated with the 6,7,8,9 directions, and the factor
of $(f_3(\wt q))^4$ in the numerator comes from the fermions
associated with the 0,3,4,5 directions. In \refb{eq3} their roles
get reversed. This gives rise to different identifications of
these expressions as
partition functions in the open string sector, as reflected in
eqs.\refb{eq1} and \refb{eq3}. Similarly in eq.\refb{eq2} the
vanishing of the integrand is due to the zero modes of fermions
associated with 6,7,8,9 directions, whereas in eq.\refb{eq4} it
is due to the zero modes of fermions associated with 0,3,4,5
directions. This again leads to different identification of these
expressions in the open string channel.
\item Since $g$ reverses the sign of $k^9$ but leaves $k^0$
invariant, only states with $k^9=0$ contribute to the trace in
the open string sector. 
The $-$ sign in eq.\refb{eq3} shows that the open
string NS sector ground state with $k^9=0$
is odd under $(-1)^F\cdot g$, and
hence is even under $g$ since it is already known to be odd under
$(-1)^F$. (The same result is reflected in the + sign in
eq.\refb{eq1}.) 
More generaly $g$ acts on a tachyonic open string state
with arbitrary momentum $(k^0,k^9)$ as
\be \label{ex1}
|k^0, k^9\ra \to |k^0, -k^9\ra\, .
\ee
\end{enumerate}

Let us now define:
\be \label{ep24}
|D1,+\ra = {1\over 2} (|U\ra_{NSNS} + |U\ra_{RR} +|T\ra_{NSNS}
+|T\ra_{RR})\, ,
\ee
\be \label{ep25}
|D1,-\ra = {1\over 2} (|U\ra_{NSNS} + |U\ra_{RR} -|T\ra_{NSNS}
-|T\ra_{RR})\, ,
\ee
\be \label{ep26}
|\wb{D1},+\ra = {1\over 2} (|U\ra_{NSNS} - |U\ra_{RR} -|T\ra_{NSNS}
+|T\ra_{RR})\, ,
\ee
\be \label{ep27}
|\wb{D1},-\ra = {1\over 2} (|U\ra_{NSNS} - |U\ra_{RR} +|T\ra_{NSNS}
-|T\ra_{RR})\, .
\ee
Using eqs.\refb{e14a}-\refb{e14d}, \refb{eq1}-\refb{eq4} we now get
\be \label{er1}
\int_0^\infty dl \,  \la D1,+|e^{-lH_c}|D1,+\ra =\int_0^\infty
{dt\over 2t} Tr_{NS-R}\Big( e^{-2t H_o} {1+(-1)^F\over 2}
{1+g\over 2}\Big)\, ,
\ee
where $Tr_{NS-R}$ denotes the difference between the traces over
the NS and R sector states. This expression is identical to the
one in \refb{ep10}, showing that the $|D1,+\ra$ state as defined
above does represent a consistent boundary state in this orbifold
theory. Similarly one can show that each of the other three
states defined in \refb{ep25}-\refb{ep27} also represent
consistent boundary states. 

The interpretation of these four boundary
states can be read out from their
expressions. The sign of $|U\ra_{RR}$ determines the sign of the
charge carried by the state under the untwisted sector RR gauge
field.  Thus if this sign is positive then the state represents a
D-string, whereas if it is negative then the state represents an
anti-D-string. On the other hand the sign of $|T\ra_{RR}$
determines the charge carried by the state under the twisted
sector RR gauge field living on the orbifold plane. 
Thus the states given in
eqs.\refb{ep24}-\refb{ep27} represent the configurations shown in
\ref{f3}(a), \ref{f3}(c), \ref{f3}(d) and \ref{f3}(b)
respectively.

\subsection{Superposition of D-strings on Orbifold} \label{s23}

We shall now consider the superposition of the D-string
configurations given in Fig.\ref{f3}(a) and \ref{f3}(d) as shown
in Fig.\ref{f4}. The corresponding boundary state is given by:
\be \label{ekk1}
|D1,+;\wb{D1},+\ra = |D1,+\ra + |\wb{D1},+\ra\, .
\ee
Thus the amplitude for the emission and reabsorption of a closed
string from this system is given by:
\ben \label{ekk2}
&& \int_0^\infty dl \,  \la
D1,+;\wb{D1},+|e^{-lH_c}|D1,+;\wb{D1},+\ra \nonumber \\
&=& \int_0^\infty dl \,  \big[ 
\la D1,+ | e^{-lH_c} | D1,+\ra +
\la \wb{D1},+ | e^{-lH_c} | \wb{D1},+\ra \nonumber \\
&& \qquad \qquad +
\la {D1},+ | e^{-lH_c} | \wb{D1},+\ra +
\la \wb{D1},+ | e^{-lH_c} | {D1},+\ra \big]\, .
\een
Each of the first two terms are given by eq.\refb{er1} and can be
interpreted as the partition function for open string with both
ends lying on the D1-brane or both ends lying on the
$\wb{D1}$-brane. On the other hand each of the last two terms
gives: 
\be \label{er3}
\int_0^\infty {dt\over 2t} Tr_{NS-R} \Big( e^{-2tH_o} {1 -
(-1)^F\over 2} {1 - g\over 2}\Big) \, .
\ee
These can be regarded as the partition functions of open string
with one end lying on the D1-brane and the other end lying on the
$\wb{D1}$-brane. The $-$ sign in front of the $(-1)^F$ term shows
that these states have opposite GSO projection, and the $-$ sign
in front of $g$ shows that these states also have opposite $g$
projection compared to open strings with both ends lying on the
same brane. Due to the opposite $(-1)^F$ projection,
the tachyon from the NS sector ground state survives. But
comparison with
eq.\refb{ex1} shows that the action of $g$ on the ground state
now takes the form:
\be \label{ex2}
|k^0, k^9\ra \to -|k^0, -k^9\ra\, .
\ee
Thus the $k^9=0$ state is projected out and the $g$ invariant
tachyonic state takes the form:
\be \label{exyz1}
|k^0,k^9\ra - |k^0, -k^9\ra\, .
\ee
If $T(x^0,x^9)$ denotes
the tachyon field on the D-string in the position space
representation, then \refb{exyz1} translates to the following
boundary condition on the tachyon field:
\be \label{ex3}
T(x^0,x^9) = 0 \qquad \hbox{at} \qquad x^9=0\, .
\ee
Thus the minimal energy configuration of this system after
tachyon condensation is expected to be described by a tachyon
field configuration of the form described in Fig.\ref{f8}.

\sectiono{D-strings in Type IIB on $(R^{8,1}\times S^1)/Z_2$}
\label{s3}

In this section we shall analyse D-string on an orbifold
$(R^{8,1}\times S^1)/Z_2$, with the D-string stretched along $S^1$,
and the $Z_2$ transformation acting by the simultaneous action of
$(-1)^{F_L}$ and the reversal of sign of the coordinate of $S^1$
and three of the coordinates of $R^{8,1}$. We begin in subsection
\ref{s31} by constructing the boundary state describing D-string
stretched along $S^1$ before the $Z_2$ modding. In subsection
\ref{s32} we discuss the result of taking the $Z_2$ orbifold of
this configuration, and construct the boundary states
corresponding to all the eight configurations displayed in
Fig.\ref{f5}. In subsection \ref{s33} we consider appropriate
superposition of these states to get the configurations of
Fig.\ref{f6} and \ref{f7}, and find in each case the critical
radius at which the tachyon becomes massless. 

\subsection{D-string on $R^{8,1}\times S^1$} \label{s31}

The analysis in this section is similar to that in section
\ref{s21}. We take the D-string to lie along the $x^9$ direction
and take this direction to be compact with period $2\pi R$. The
main difference from the analysis of section \ref{s21} arises due
to the fact that in this case there is a new quantum number, $-$
the winding number along $S^1$. We shall refer to the integer
labelling the winding number as $w_9$. We now define new coherent
states:
\ben \label{es1}
|k,w_9,\eta\ra_{NSNS\atop RR} 
&=& \exp\Big( \sum_{n=1}^\infty \big[
-{1\over n}\sum_{\mu = 0,9}\alpha^\mu_{-n} \wt \alpha^\mu_{-n}
+{1\over n}\sum_{\mu =3}^8\alpha^\mu_{-n} \wt
\alpha^\mu_{-n}\big] \nonumber \\
&& + i\eta \sum_{r>0} \big[ 
-\sum_{\mu=0,9} \psi^\mu_{-r}\wt\psi^\mu_{-r}
+\sum_{\mu=3}^8 \psi^\mu_{-r}\wt\psi^\mu_{-r}\big]\Big)
|k,w_9,\eta\ra_{NSNS\atop RR}^{(0)}\, ,
\een
where the vacua $|k,w_9,\eta\ra^{(0)}$ in the NSNS and 
RR sectors are defined in a manner analogous to 
$|k,\eta\ra^{(0)}$ as in section \ref{s21}.
We now take linear combinations of these states of the form:
\be \label{es2}
|\theta, \eta\ra_{NSNS} = \NN \sum_{w_9} e^{i\theta w_9} \int
\Big(\prod_{\mu=1}^8 dk^\mu\Big) | k,w_9, \eta\ra_{NSNS}\, ,
\ee
and, 
\be \label{es3}
|\theta, \eta\ra_{RR} = 4i\NN \sum_{w_9} e^{i\theta w_9} \int
\Big(\prod_{\mu=1}^8 dk^\mu\Big) | k,w_9, \eta\ra_{RR}\, ,
\ee
$\NN$ being the same normalization constant defined in
\refb{e16}. We now define
\be \label{es3a}
|\theta,U\ra_{NSNS} = {1\over\sqrt 2} (|\theta,+\ra_{NSNS} -
|\theta, -\ra_{NSNS})\, ,
\ee
\be \label{es4}
|\theta,U\ra_{RR} = {1\over\sqrt 2} (|\theta,+\ra_{RR} +
|\theta, -\ra_{RR})\, ,
\ee
\be \label{es5}
|\theta, D1\ra = {1\over\sqrt 2} (|\theta, U\ra_{NSNS} +|\theta,
U\ra_{RR})\, ,
\ee
\be \label{es6}
|\theta, \wb{D1}\ra = {1\over\sqrt 2} (|\theta, U\ra_{NSNS} 
-|\theta, U\ra_{RR})\, .
\ee
In order to find a physical interpretation of the parameter
$\theta$, let us analyse the term
\be \label{es7}
\int_0^\infty dl \,  \la \theta, D1| e^{-lH_c}|\theta', D1\ra\, ,
\ee
which should give us the partition function of the open strings
with one leg on the string described by the state $|\theta,D1\ra$
and the other leg on the string described by the state
$|\theta',D1\ra$. $H_c$ now has an additional term $\pi R^2
w_9^2$ compared to \refb{e10}.
Upon evaluation following standard method, we
find that the integrand acquires an extra factor of
\be \label{es8}
\sum_{w_9} e^{iw_9 (\theta' -\theta) -l\pi R^2 w_9^2}\, ,
\ee
compared to the integrand of \refb{ep1}. Using a Poisson
resummation, we can rewrite \refb{es8} as
\be \label{es8a}
{1\over R\sqrt l} \sum_m e^{-2t\pi {1\over R^2} \{ m -
{\theta-\theta'\over 2\pi}\}^2}\, ,
\ee
where $t=(2l)^{-1}$ as usual.
In this equation the sum over $m$ runs over integers. Since
$R\sqrt l$ can be rewritten as
\be \label{es9}
R\int dk_9 e^{-2t\pi k_9^2}\, ,
\ee
we see that the effect of multiplication by
\refb{es8a} is to replace \refb{es9} in the trace over open
string states by,
\be \label{es10}
\sum_{k_9= {1\over R} (m -{\theta-\theta'\over 2\pi})} e^{-2t\pi
k_9^2}\, .
\ee
The reason for replacing the integral over $k_9$ by
a discrete sum can
be attributed to the fact that the 9th direction is compact.
However, note that instead of the usual quantization of $k_9$ 
in units of
$1/R$, here we have an additional constant shift by
$(\theta-\theta')/2\pi R$. Thus the wave-functions of these states,
instead of being periodic along $x^9$, pick up a phase of
$\exp(-i(\theta-\theta'))$ as $x^9\to x^9+2\pi R$. 
This can be attributed to the
presence of Wilson
lines $\theta$ and $\theta'$ associated with the U(1)
gauge fields living on the two D-strings. Thus the parameter
$\theta$ measures the Wilson line associated with the U(1) gauge
field living on the D-string.

\subsection{D-strings on $(R^{8,1}\times S^1)/Z_2$} \label{s32}

We now take the orbifold of the configuration described in the
previous section by the $Z_2$ group generated by $(-1)^{F_L}$
accompanied by the reversal of sign of $x^6$, $x^7$, $x^8$ and
$x^9$. For our purpose it will be convenient to start with a
circle of radius $2R$, and define two $Z_2$
transformations $g_1$ and $g_2$ as follows. $g_1$ correponds to
the transformation 
\be \label{es12}
x^6\to -x^6, \quad x^7\to -x^7, \quad x^8\to - x^8, \quad x^9\to
-x^9\, ,
\ee
accompanied by the transformation $(-1)^{F_L}$. On the other hand
$g_2$ corresponds to the transformation
\be \label{es13}
x^6\to -x^6, \quad x^7\to -x^7, \quad x^8\to - x^8, \quad x^9\to
2\pi R-x^9\, ,
\ee
accompanied by the transformation $(-1)^{F_L}$. Then
$g_2 g_1$ generates
\be \label{es14}
g_2g_1: \qquad x^9\to x^9 + 2\pi R\, .
\ee
Thus modding out by the group generated by $g_1$ and $g_2$ gives
us back a circle of radius $R$.

With this convention, the $w_9$ even states appearing in
\refb{es2}, \refb{es3} may be regarded as untwisted sector
states, whereas the $w_9$ odd states belong to the sector twisted
by $g_1g_2$.
Since the transformation $g_1$ reverses the sign of $w_9$ in
$|k,w_9,\eta\ra$, we see from \refb{es2}, \refb{es3} that only
the $|0,\eta\ra_{NSNS\atop RR}$ and $|\pi, \eta\ra_{NSNS\atop
RR}$ states are invariant under the orbifold projection. 
As in section
\ref{s22}, we need to add twisted sector coherent states to
\refb{es5} and \refb{es6} in order that the amplitude for
emission and reabsorption of a closed string can be reexpressed
as open string partition function with appropriate 
projection. Since there are two sets of twisted sector states,
one living at $x^9=0$ and twisted by $g_1$,
and the other living at $x^9=\pi R$ and twisted by $g_2$, we now
have two sets of states of the kind described in \refb{ep22}, 
\refb{ep23}:
$|T_1\ra_{NSNS\atop RR}$ and $|T_2\ra_{NSNS\atop RR}$ defined
through eqs.\refb{ep11}-\refb{ep23}. Our task now is to find
appropriate linear combinations of these states so that the
amplitude for the emission and reabsorption of a closed string
can be reexpressed as an open string partition function with
appropriate projections. An additional consistency requirement
comes from the fact that if the two ends of the open string are
on D-strings carrying the same Wilson line (0 or $\pi$), then 
due to eq.\refb{es10} the wave-function is periodic in $x^9$, 
and hence by 
eq.\refb{es14} the action of $g_1$ and $g_2$ on the state must be
identical. On the other hand if one end of the open string is on
a D-string without Wilson line and the other end is on a D-string
with Wilson line $\pi$, then the wave-function is anti-periodic
in $x^9$ and hence the action of $g_1$ and $g_2$ on the state
must differ by a sign.

We find eight consistent boundary states satisfying these
conditions. They are labelled by three quantum numbers, the
Wilson line $\theta$ which can take values 0 or $\pi$, and two
other numbers $\eps_1$ and $\eps_2$ each of which can take values
$\pm 1$. The boundary states are given by:
\ben \label{es17}
|\theta,\eps_1,\eps_2\ra &=& {1\over 2} (|\theta,U\ra_{NSNS}
+ \eps_1 |\theta,U\ra_{RR}) + {1\over 2\sqrt 2} \eps_2
(|T_1\ra_{NSNS} + \eps_1 |T_1\ra_{RR}) \nonumber \\
&& \qquad \qquad + {1\over 2\sqrt 2} e^{i\theta} \eps_2 
(|T_2\ra_{NSNS} + \eps_1 |T_2\ra_{RR})
\een
The various parameters have the following interpretation:
\begin{enumerate}
\item $\eps_1$ determines the orientation of the D-string. We
shall choose the convention that positive $\epsilon_1$ means that
the string is oriented towards left in Fig.\ref{f5}, {\it i.e.}
from $x^9=\pi R$ towards $x^9=0$.
\item $\eps_1\eps_2$ denotes the charge carried by the twisted
sector RR gauge field living at $x^9=0$.
\item $\eps_1\eps_2 e^{i\theta}$ denotes the charge carried by
the twisted sector RR gauge field living at $x^9=\pi R$.
\end{enumerate}
Thus the eight different configurations displayed in Fig.\ref{f5} 
are described by the following sets of values of
$(\theta,\eps_1,\eps_2)$:\footnote{These configurations are
T-dual to the fractionally charged D0-branes discussed in
\cite{DOUMOO,POLC,JON,DTH}. An analogous phenomenon was discussed
in \cite{LUST}.}
\ben \label{es17a}
(a)&:& \qquad (\pi,+,+) \cr
(b)&:& \qquad (0,+,-) \cr
(c)&:& \qquad (0,+,+) \cr
(d)&:& \qquad (\pi,+,-) \cr
(e)&:& \qquad (\pi,-,-) \cr
(f)&:& \qquad (0,-,+) \cr
(g)&:& \qquad (0,-,-) \cr
(h)&:& \qquad (\pi,-,+) \, .
\een
Using the definition \refb{es17} of the state
$|\theta,\eps_1,\eps_2\ra$ one can easily verify that
\ben \label{es18}
&& \int_0^\infty dl \,  \la \theta', \eps_1', \eps_2'| e^{-l H_c}|
\theta,\eps_1, \eps_2\ra  \nonumber \\
&=& \int_0^\infty {dt\over 2t} Tr_{NS-R}
\Big\{ e^{-2t H_o} {1 + \eps_1\eps_1'(-1)^F\over 2} { 1 +
\eps_2\eps_2' g_1\over 2} {1 + \eps_2\eps_2'
e^{i(\theta-\theta')} g_2\over 2}\Big\}\, .
\een
{}From this we see that $g_1$ and $g_2$ have the same (opposite)
projection if $(\theta-\theta')$ equals zero ($\pi$), as required.
The sum over $k_9$ in the trace runs over $m/2R$ for all integers
$m$.

\subsection{Superpositions of D-strings on $(R^{8,1}\times
S^1)/Z_2$} \label{s33}

Let us now consider the superposition of the configurations given
in Figs.\ref{f5}(a) and \ref{f5}(g). Our object of interest is the
partition function of the open string states with two legs lying
on the two different strings. Using eqs.\refb{es17a} and
\refb{es18} we see that this contribution is given by:
\be \label{es21}
\int_0^\infty {dt\over 2t} Tr_{NS-R}
\Big\{ e^{-2t H_o} {1 - (-1)^F\over 2} { 1 -
g_1\over 2} {1 +  g_2\over 2}\Big\}\, .
\ee
Since this gives projection onto states with $(-1)^F=-1$, the
`tachyon' of mass$^2$ $-(2\alpha')^{-1}$
from the NS sector ground state survives. However since
we need to project onto $g_1g_2=-1$ states, the momentum $k_9$
is quantized as $(m+{1\over 2})R^{-1}$ with $m$ integer,
and hence the $k^9=0$ mode is
projected out. Thus the effective mass of the lowest mode,
carrying $k^9=\pm{1\over 2R}$ is given by
\be \label{es23}
m^2 = (k^9)^2 - {1\over 2\alpha'} = {1\over 4R^2} - {1\over
2\alpha'}\, .
\ee
{}From this we see that the tachyonic mode disappears for
\be \label{es24}
R \le \sqrt{\alpha'\over 2} \equiv R_c\, .
\ee
Thus if we identify the total mass of the D-string anti-D-string
pair at this value of $R$ as $m_{++}$, we get
\be \label{erep1}
m_{++} = 2\pi T_D R_c = (\sqrt{2\alpha'}g)^{-1} \, .
\ee

Let us also consider the superposition of Figs.\ref{f5}(c) and
\ref{f5}(g) as shown in Fig.\ref{f7}. Using eqs.\refb{es17a} and
\refb{es18} we see that the partition function of the open string
with two ends lying on the two different D-strings is now given
by
\be \label{es26}
\int_0^\infty {dt\over 2t} Tr_{NS-R}
\Big\{ e^{-2t H_o} {1 - (-1)^F\over 2} { 1 -
g_1\over 2} {1 -  g_2\over 2}\Big\}\, .
\ee
Since this gives projection onto states with $(-1)^F=-1$, the
`tachyon' of mass$^2$ $-(2\alpha')^{-1}$
from the NS sector ground state survives as in the previous case. 
But since we now need to
project onto $g_1g_2=1$ states, the momentum is quantized as
$m/R$ with $m$ integer.
However, the $k_9=0$ mode is still projected out,
as the states are required to be odd under the action of $g_1$
and $g_2$ which transform $k_9$ to $-k_9$.  
Thus the effective mass of the lowest mode,
carrying $k^9=\pm{1\over R}$ is given by
\be \label{es31}
m^2 = (k^9)^2 - {1\over 2\alpha'} = {1\over R^2} - {1\over
2\alpha'}\, .
\ee
{}From this we see that the tachyonic mode disappears for
\be \label{es32}
R \le \sqrt{2\alpha'} \equiv R_c'\, .
\ee
At this critical radius, we expect the mass of the D-string 
anti-D-string pair to be equal to twice the 
mass of a ++ state living
on the orbifold plane. This gives,
\be \label{erep22}
2 m_{++} = 2\pi T_D R_c' = 2(\sqrt{2\alpha'}g)^{-1} \, .
\ee
This is the same as eq.\refb{erep1}.

Acknowledgement: I wish to thank O. Bergman and M. Gaberdiel for 
useful correspondence.

\appendix

\sectiono{Notations and Normalization Conventions} \label{sa}

We work with $\alpha'=1$, {\it i.e.} string tension equal to
$1/2\pi$. By a double Wick rotation we make the coordinate $x^0$
space-like and the coordinate $x^1$ time-like, so that we can
formulate the theory in the light-cone gauge with $x^1\pm x^2$ as
the light-cone coordinates. 
The light-cone gauge action for closed string is then
given by:
\ben \label{ea1}
 &&{1\over 4\pi} \int d\tau\int_0^1 d\sigma \sum_{\mu=0, 3\ldots 9}
\Big(\p_\tau X^\mu \p_\tau X^\mu -\p_\sigma X^\mu \p_\sigma X^\mu
\nonumber \\
&& + i \psi^\mu (\p_\tau + \p_\sigma) \psi^\mu 
+ i \wt\psi^\mu (\p_\tau - \p_\sigma) \wt\psi^\mu \Big)\, ,
\een
where $X^\mu$, $\psi^\mu$ and $\wt\psi^\mu$
are periodic functions of $\sigma$ with period 1. The equations
of motion derived from this action give the following mode
expansion for the fields;
\ben \label{ea2}
X^\mu(\tau,\sigma) &=& x^\mu + 2\pi p^\mu\tau + {i\over \sqrt 2}
\sum_{n\ne 0} {1\over n} (\alpha^\mu_n e^{-2\pi i n(\tau
-\sigma)} + \wt\alpha^\mu_n e^{-2\pi i n(\tau+\sigma)})\, ,
\nonumber \\
\psi^\mu(\tau, \sigma) &=& \sqrt{2\pi}
\sum_r \psi^\mu_r e^{-2\pi i r(\tau -
\sigma)}\, , \nonumber \\
\wt \psi^\mu(\tau, \sigma) &=& \sqrt{2\pi}
\sum_r \wt\psi^\mu_r 
e^{-2\pi i r(\tau + \sigma)}\, , 
\een
for $\mu=0, 3,4\ldots 9$. The sum over $n$ runs over integers,
whereas the sum over $r$ runs over integers and 
(integers + ${1\over 2}$) in the Ramond (R) and Neveu-Schwarz
(NS) sector respectively. Besides these oscillator modes, we also
need to include the zero modes $x^1,x^2,p^1,p^2$ in our list of
dynamical variables. The commutation relations satisfied by these
various oscillators are as follows:
\ben \label{ea3}
&&[\alpha^\mu_m, \alpha^\nu_n] = [\wt\alpha^\mu_m, \wt\alpha^\nu_n]
= m \delta_{\mu\nu}\delta_{m+n,0}, 
\qquad \hbox{for} \quad \mu,\nu =0,3,4\ldots 9\, ,
\nonumber \\
&& [x^\mu, p^\nu] = i\eta^{\mu\nu} \qquad \hbox{for} \quad
0\le\mu,\nu \le 9, \nonumber \\
&& \{\psi_r^\mu, \psi_s^\nu\} = \{\wt\psi_r^\mu, \wt\psi_s^\nu\}
= \delta_{\mu\nu}\delta_{r+s,0}\, , \qquad \hbox{for} \quad
\mu,\nu=0, 3,4\ldots 9\, .
\een
Here $\eta^{\mu\nu}$ is a diagonal matrix with $\eta^{11}$ equal
to $-1$ and all the other diagonal entries equal to $+1$. All other
(anti-) commutators vanish.

Physical states are required to be invariant under the GSO
operators $(-1)^F$ and $(-1)^{\wt F}$, which change the sign of
all the world-sheet fermion fields $\psi^\mu$ and $\wt\psi^\mu$
respectively. The ground state of the NS sector is taken to be
odd under these operations. Due to the presence of the fermion
zero modes the Ramond sector ground state is degenerate. We
choose our convention such that the Ramond sector ground states
$|\Omega\ra_{L,R}$ satisfying
\be \label{ea4}
\prod_{\mu=0,3,4\ldots 9} (\sqrt 2 \psi_0^\mu)
|\Omega\ra_R = |\Omega\ra_R \, , \qquad
\prod_{\mu=0,3,4\ldots 9} (\sqrt 2 \wt\psi_0^\mu)
|\Omega\ra_L = |\Omega\ra_L\, ,
\ee
are even under $(-1)^F$ and $(-1)^{\wt F}$ respectively.

An open string with two ends lying on a D-string along the $x^9$
direction is described by the same action as 
\refb{ea1}. However, instead of being periodic in $\sigma$, $X^\mu$'s
now satisfy Neumann (Dirichlet) boundary condition for $\mu = 0,9$ 
($\mu = 3,4, .. 8$). The mode expansions are given by:
\ben \label{ea5}
X^\mu(\tau,\sigma) &=& x^\mu + 2\pi p^\mu\tau + {i \sqrt 2}
\sum_{n\ne 0} {1\over n} \alpha^\mu_n e^{-\pi i n\tau}
\cos n\pi\sigma \qquad \hbox{for} \quad \mu=0,9\, ,
\nonumber \\
 &=&  {i \sqrt 2}
\sum_{n\ne 0} {1\over n} \alpha^\mu_n e^{-\pi i n\tau}
\sin n\pi\sigma \qquad \hbox{for} \quad \mu=3,4\ldots 8\, ,
\nonumber \\
\psi^\mu(\tau, \sigma) &=& \sqrt{\pi}
\sum_r \psi^\mu_r e^{-\pi i r(\tau -\sigma)}\qquad \hbox{for}
\qquad \mu = 
0,3,4\ldots 9
\, , \nonumber \\
\wt \psi^\mu(\tau, \sigma) &=& \sqrt{\pi}
\sum_r \psi^\mu_r 
e^{-\pi i r(\tau + \sigma)} \qquad \hbox{for} \quad \mu=0,9  \, ,
\nonumber \\ 
 &=& - \sqrt{\pi}
\sum_r \psi^\mu_r 
e^{-\pi i r(\tau + \sigma)} \qquad \hbox{for} \quad \mu=3,4\ldots 8\, ,
\een
The commutation relations of these oscillators are also given 
by eq.\refb{ea3}. Physical states are required to be invariant
under the operator $(-1)^F$ which changes the sign of the
oscillators $\psi^\mu_r$. The ground state of the NS sector is
again taken to be odd under $(-1)^F$, and the ground state of the
Ramond sector satisfying
\be \label{efi1}
\prod_{\mu=0,3,4,\ldots 9} \psi_0^\mu |\Omega\ra = |\Omega\ra\, ,
\ee
is taken to be even under $(-1)^F$.


\begin{thebibliography}{99}

\bibitem{NONBP}
A. Sen, [hep-th/9803194].

\bibitem{BERKOL}
O. Bergman and B. Kol, [hep-th/9804160].

\bibitem{DBRANE}
J. Polchinski, Phys. Rev. Lett. {\bf 75} (1995) 4724
[hep-th/9510017]; \\
J. Dai, R. Leigh and J. Polchinski, Mod. Phys. Lett. {\bf A4}
(1989) 2073; \\
R. Leigh, Mod. Phys. Lett. {\bf A4} (1989) 2767; \\
J. Polchinski, Phys. Rev. {\bf D50} (1994) 6041 [hep-th/9407031].

\bibitem{ORIENT}
A. Sagnotti, `Open Strings and their Symmetry Groups', Talk at
Cargese Summer Inst., 1987; \\
G. Pradisi and A. Sagnotti, Phys. Lett. {\bf B216} (1989) 59; \\
M. Bianchi, G. Pradisi and A. Sagnotti, Nucl. Phys. {\bf B376}
(1992) 365; \\
P. Horava, Nucl. Phys. {\bf B327} (1989) 461; Phys. Lett. {\bf
B231} (1989) 251; \\
E. Gimon and J. Polchinski, Phys. Rev. {\bf D54} (1996) 1667
[hep-th/9601038].

\bibitem{DUORBI}
A. Sen, Nucl.Phys. {\bf B474} (1996) 361 [hep-th/9604070].

\bibitem{BERGAB}
O. Bergman and M. Gaberdiel, Nucl.Phys. {\bf B499} (1997) 183
[hep-th/9701137].

\bibitem{POLCAI}
J. Polchinski and Y. Cai, Nucl.Phys. {\bf B296} (1988) 91.

\bibitem{CANAYO}
C. Callan, C. Lovelace, C. Nappi and S. Yost, Nucl.Phys. {\bf
B308} (1988) 221.

\bibitem{MLI}
M. Li, Nucl.Phys. {\bf B460} (1996) 351 [hep-th/9510161].

\bibitem{OTHER}
H. Ooguri, Y. Oz and Z. Yin, Nucl.Phys. {\bf B477} (1996) 407
[hep-th/9606112]; \\
K. Becker, M.Becker, D. Morrison, H. Ooguri, Y. Oz and Z. Yin,
Nucl. Phys. {\bf B480} (1996) 225 [hep-th/9608116]; \\
M. Kato and T. Okada, Nucl. Phys. {\bf B499} (1997) 583
[hep-th/9612148]; \\
S. Stanciu, [hep-th/9708166]; \\
A. Recknagel and V. Schomerus, [hep-th/9712186]; \\
J. Fuchs and C. Schweigert, [hep-th/9712257]; \\
S. Stanciu and A. Tseytlin, [hep-th/9805006].

\bibitem{IEN}
F. Hussain, R. Iengo, C. Nunez and C. Scrucca, Phys. Lett. {\bf
B409} (1997) 101 [hep-th/9706186]; \\
M. Bertolini, R. Iengo and C. Scrucca, [hep-th/9801110]; \\
M. Bertolini, P. Fre, R. Iengo and C. Scrucca, [hep-th/9803096].

\bibitem{POLSTR}
J. Polchinski and A. Strominger, Phys. Lett. {\bf B388} (1996) 736
[hep-th/9510227].

\bibitem{NARA}
E. Gava, K. Narain and M. Sarmadi, Nucl.Phys. {\bf B504} (1997)
214 [hep-th/9704006]; \\
I. Antoniadis, E. Gava, K.Narain and T. Taylor, Nucl. Phys. {\bf
B511} (1998) 611 [hep-th/9708075].

\bibitem{SMILGA}
A. Smilga and A. Veselov, Nucl. Phys. {\bf B515} (1998) 163
[hep-th/9710123]; [hep-th/9801142].

\bibitem{BANSUS}
T. Banks and L. Susskind, [hep-th/9511194].

\bibitem{GRGUP}
M. Green and M. Gutperle, Nucl.Phys.{\bf B476} (1996) 484
[hep-th/9604091].

\bibitem{PERI}
G. Lifschytz, Phys. Lett. {\bf B388} (1996) 720 [hep-th/9604156];
\\
V. Periwal, [hep-th/9612215].

\bibitem{DOUMOO}
M. Douglas and G. Moore, [hep-th/9603167].

\bibitem{POLC}
J. Polchinski, Phys. Rev. {\bf D55} (1997) 6423 [hep-th/9606165].

\bibitem{JON}
C. Johnson and R. Myers, Phys. Rev. {\bf D55} (1997) 6382
[hep-th/9610140].

\bibitem{DTH}
D. Diaconescu, M. Douglas and J. Gomis, [hep-th/9712230].

\bibitem{LUST}
A. Karch, D. Lust and D. Smith, [hep-th/9803232].


\end{thebibliography}
\end{document}